\documentclass[reprint,superscriptaddress,aps,prd,floatfix,nofootinbib,preprintnumbers]{revtex4-2}

\usepackage[utf8]{inputenc} 
\usepackage{graphicx}
\usepackage[dvipsnames]{xcolor}

\usepackage{amsmath,amsfonts,amssymb,bm}
\usepackage{slashed}
\usepackage{hyperref}
\hypersetup{colorlinks, urlcolor=BlueViolet, citecolor=Plum, linkcolor=PineGreen, linktocpage=true}
\usepackage{orcidlink}

\begin{document}

\title{Dark Matter Induced Nucleon Decay Through the Neutron Portal}

\author{Nicole F. Bell \orcidlink{0000-0002-5805-9828}}
\email{n.bell@unimelb.edu.au}

\author{Peter Cox \orcidlink{0000-0002-6157-3430}}
\email{peter.cox@unimelb.edu.au}

\author{Jayden L. Newstead \orcidlink{0000-0002-8704-3550}}
\email{jnewstead@unimelb.edu.au}

\author{Michael B. G. Verde \orcidlink{0009-0006-5570-4517}}
\email{mverde@student.unimelb.edu.au}
\affiliation{ARC Centre of Excellence for Dark Matter Particle Physics, \\
School of Physics, The University of Melbourne, Victoria 3010, Australia}

\begin{abstract}

The neutron portal operator provides a theoretically motivated connection between the visible and dark sectors and features in several well-studied asymmetric dark matter models. This operator leads to dark matter induced nucleon decays that mimic the experimental signature of ``ordinary" nucleon decays. In this work, we reinterpret Super-Kamiokande nucleon decay searches for $n \rightarrow \pi^0 \nu$ and $p \rightarrow \pi^+ \nu$ to constrain dark matter induced nucleon decays. For GeV-scale dark matter, we obtain lower bounds of $\mathcal{O}(1~\rm{TeV})$ on the scale of the effective neutron portal operator. We also discuss the prospects for future searches at Hyper-Kamiokande and highlight the importance of a dedicated experimental analysis with reduced systematic uncertainties. 

\end{abstract}

\maketitle

\section{Introduction}
\label{Sec:Intro}

One of the most promising avenues to understand the nature of dark matter (DM) is to directly detect it in terrestrial experiments. Direct detection experiments have primarily focused on searches for the elastic scattering of DM with Standard Model (SM) particles, $\text{DM} + \text{SM} \to \text{DM} + \text{SM}$, where the relevant SM particles are either atomic nuclei or electrons. However, within the broad landscape of DM candidates, there exists a much wider range of possible experimental signatures, including those in which the DM and/or SM particles differ between the initial and final states\footnote{Examples of such processes include the absorption of fermionic DM~\cite{Ando:2010ye,Dror:2019onn,PandaX:2022osq} and inelastic scattering within multi-component dark sectors~\cite{Tucker-Smith:2001myb,Barello:2014uda,LZ:2023lvz}.}, i.e. $\text{DM} + \text{SM} \to \text{DM}' +\text{SM}'$. 

A particularly well-motivated class of DM models are those that lie within the paradigm of asymmetric DM (ADM) (for a review see \cite{Davoudiasl:2012uw,Petraki:2013wwa,Zurek:2013wia}). These models are motivated by the apparent coincidence between the energy densities of baryonic and dark matter ($\Omega_\text{DM} \simeq 5 \Omega_b$) and seek to explain this coincidence by connecting the matter--anti-matter asymmetry in the visible sector with a particle--anti-particle asymmetry in the dark sector.\footnote{A complete explanation of the coincidence also requires a connection between the DM mass and the proton mass.} Perhaps the most interesting subclass of ADM models are those in which the DM itself carries (a generalised) baryon number~\cite{Dodelson:1989cq}. These models feature a distinctive experimental signature: the DM induced destruction/decay of baryons~\cite{Davoudiasl:2010am}.

Experiments searching for proton decay and exotic neutron decays are potentially also highly sensitive to DM-induced nucleon decays (INDs). In this work, we reinterpret data from the Super Kamiokande (Super-K) searches for $p \to \pi^+ \nu$ and $n \to \pi^0 \nu$~\cite{Super-Kamiokande:2013rwg} to constrain the DM-induced processes $\text{DM} + p \to \pi^+ + \text{DM}'$ and $\text{DM} + n \to \pi^0 + \text{DM}'$, where $\text{DM}'$ represents a second dark sector particle. 

The low-energy effective theory of ADM includes operators that provide the connection between the SM and dark sectors. In this work, we derive new constraints on the dimension-7 neutron portal operator, $(u_R d_R)(d_R \Psi_R) \Phi$, where $\Psi,\Phi$ are dark sector fields. This operator arises in well-studied UV completions of ADM such as Hylogenesis~\cite{Davoudiasl:2010am}. In fact, the IND processes we consider were first studied in the context of this model in Ref.~\cite{Davoudiasl:2011fj}, in its supersymmetric extension in~\cite{Blinov:2012hq}, and most recently within an effective field theory (EFT) framework in~\cite{Liang:2023yta}.

IND processes can also arise in other types of models. The $p \to \pi^+$ and $n \to \pi^0$ INDs occur within Mesogenesis models of baryogenesis~\cite{Elor:2018twp}, although via a different operator. The sensitivity of future experiments to INDs in Mesogenesis was investigated in Ref.~\cite{Berger:2023ccd}. IND is also predicted in certain models seeking to explain the neutron lifetime anomaly~\cite{Jin:2018moh,Keung:2019wpw,Fornal:2020gto,Strumia:2021ybk,Liang:2023yta}. IND processes with leptonic final states have also been studied~\cite{Huang:2013xfa,Demidov:2015bea,Ema:2024wqr}, and the complementarity with di-nucleon decays was recently explored in Ref.~\cite{ThomasArun:2025gvk}.

In this paper, we obtain the first robust limits on the neutron portal operator from the IND processes $N + \Phi \to \bar\Psi + \pi$ and $ N + \Psi \to \Phi^\dagger + \pi$, where $N=(p,n)$, by performing a detailed reinterpretation of the Super-K analysis in Ref.~\cite{Super-Kamiokande:2013rwg}.\footnote{Bounds on the neutron portal operator from $n + \Psi \to \Phi^{\dagger} + \pi^0$ were previously derived in Ref.~\cite{Liang:2023yta} for a restricted range of $\Phi$ and $\Psi$ masses motivated by the neutron decay anomaly. In this work, we consider all IND processes and perform a detailed statistical analysis with a careful treatment of signal efficiencies.
}
The kinematics of the IND processes differ from the corresponding ``ordinary" nucleon decays, with the final-state pion energy determined by the masses of the two dark sector states. We obtain lower bounds of $0.5-2$\,TeV on the EFT scale of the dim-7 neutron portal operator, depending on the DM mass, DM spin and mass splitting. We also investigate the prospects for future IND searches at Hyper-Kamiokande (Hyper-K), highlighting both the additional parameter space that could be probed by a dedicated experimental search and the need for further improvements in the modelling of nuclear effects to reduce systematic uncertainties.

\section{Induced Nucleon Decay}
\label{Sec:EFT}

In this paper, we focus on the well-motivated dimension-7 neutron portal operator given by (in two-component notation)\footnote{The operator with Lorentz structure $\epsilon_{\alpha \beta \gamma}(u_R^{\alpha} \Psi_R) (d_R^{\beta} d_R^{\gamma})$ vanishes due to the interplay of the spinor structure and anti-symmetric colour indices on the $d$-quarks.}
\begin{equation}
    \mathcal{L}_{int} = \frac{1}{\Lambda^3}\epsilon_{\alpha \beta \gamma}\Phi (u_R^{\alpha} d_R^{\beta}) (d_R^{\gamma}\Psi_R) \, ,
    \label{Eq:Neutron_Port_Op}
\end{equation}
where $\{ \alpha, \beta, \gamma \}$ denote colour indices that will henceforth be suppressed. The two new dark sector particles are a complex scalar, $\Phi$, and a Dirac fermion, $\Psi$. The interaction strength between the visible and dark sectors is parameterised by the scale $\Lambda$. There is a second dim-7 neutron portal operator, obtained with the replacement $\Psi_R \to \Psi_L^\dagger$. Both operators have identical IND signatures and our analysis is applicable to either operator. We remain agnostic to the UV completion of the operator and separately consider the cases where either $\Phi$ or $\Psi$ constitute all of the the DM.

\begin{figure}
    \centering
    \includegraphics[width=\linewidth]{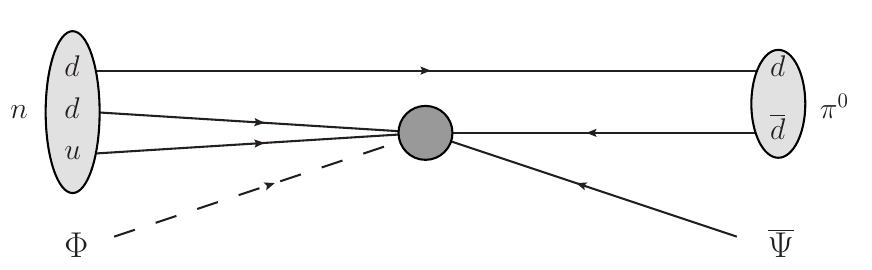}
    \caption{An example neutron-$\Phi$ IND process resulting from the dimension-7 neutron portal operator of Eq.~\eqref{Eq:Neutron_Port_Op}.}
    \label{Fig:IND_diagram}
\end{figure}

The operator in Eq.~\eqref{Eq:Neutron_Port_Op} allows interactions of the form
\begin{align}
    N + \Phi \rightarrow \bar{\Psi} + \pi \, , \label{Eq:Phi_IND} \\
    N + \Psi \rightarrow \Phi^{\dagger} + \pi \, , \label{Eq:Psi_IND}
\end{align}
where $N$ is a nucleon. A representative Feynman diagram for these processes is shown in Fig.~\ref{Fig:IND_diagram}. These processes are known as induced nucleon decays as they can mimic the experimental signature of the nucleon decay $N \rightarrow \pi \nu$. We refer to the IND processes in Eq.~\eqref{Eq:Phi_IND} and Eq.~\eqref{Eq:Psi_IND} as nucleon-$\Phi$ and nucleon-$\Psi$ IND, respectively. In this work, we only consider the operator containing first generation quark fields; operators with second and third generation quarks can lead to IND processes with different final states, such as $N + \Phi \rightarrow \bar{\Psi} + K$~\cite{Davoudiasl:2011fj}.

For the IND processes to be kinematically permitted, the dark sector particles must satisfy certain mass requirements. The nucleon-$\Phi$ IND process, for example, requires $m_N + m_{\Phi} > m_{\Psi} + m_{\pi}$. Additionally, to ensure that the DM is stable, the decay channel $\Phi \rightarrow \bar{\Psi} n \rightarrow \bar{\Psi}p\bar{\nu}_e e^-$ must be kinematically forbidden, which requires $m_{\Phi} < m_{\Psi} + m_p +m_e$. Analogous mass constraints apply for the nucleon-$\Psi$ process, where $\Psi$ is the DM. Putting the pieces together, the mass requirements for the nucleon-$\Phi$ IND process are
\begin{equation}
    m_{\pi} - m_N < \Delta m < m_p + m_e \, , 
    \label{Eq:Phi_Mass_Splitting_Requirements}
\end{equation}
while for the nucleon-$\Psi$ IND process
\begin{equation}
    -(m_p + m_e) <\Delta m < m_N - m_{\pi} \, ,
    \label{Eq:Psi_Mass_Splitting_Requirements}
\end{equation}
where $\Delta m \equiv m_{\Phi} - m_{\Psi}$ is the mass splitting between the two dark sector particles and $m_N$, $m_{\pi}$, $m_p$ and $m_e$ are the masses of the initial-state nucleon, the outgoing pion, the proton, and the electron, respectively. For low DM masses, the proton decay channel $p \rightarrow \pi^+ \Phi \Psi$ is kinematically accessible, which leads to much stronger bounds.

In specific ADM models, e.g., Hylogenesis~\cite{Davoudiasl:2010am}, there is an additional restriction on the masses to ensure that the correct DM relic abundance is obtained. We do not impose any additional constraints in our analysis, thus our results can be reinterpreted in the context of any model that generates the neutron portal operator, Eq.~\eqref{Eq:Neutron_Port_Op}, at low energies. In appendix \ref{Sec:Appdx_Hylogenesis}, we provide constraints on the Hylogenesis scenario of Ref.~\cite{Davoudiasl:2011fj}.

\subsection{Induced nucleon decay cross section}
\label{Sec:IND_cross_section}

In this section, we calculate the cross sections for the IND processes. For proton-$\Phi$ IND interactions, the matrix element is
\begin{align}
    & \left(i \mathcal{M}\right)_{\Phi p} = \frac{1}{\Lambda^3} \bar{u} (p_{\Psi}) \langle \pi^+| (u_R d_R) d_R | p \rangle \label{Eq:nPhi_Matrix_Element} \\
    & =\frac{1}{\Lambda^3} \bar{u} (p_{\Psi}) P_R \left( W^{RR}_0(q^2) + \frac{\slashed{q}}{m_p} W^{RR}_1(q^2)\right) u (p_p) \notag \,,
\end{align}
where $P_R$ is the right-handed projection operator, and $\bar{u}$ and $u$ are the standard 4-component Dirac spinors for $\Psi$ and the proton, respectively. The matrix element for the proton-$\Psi$ IND processes is obtained by substituting $\bar{u} \rightarrow \bar{v}$ in Eq.~\eqref{Eq:nPhi_Matrix_Element}. The form factors $W_0(q^2)$ and $W_1(q^2)$ parametrise the proton-to-pion matrix element as a function of the momentum transfer of the process, $q = p_p - p_{\pi^+}$. 

The matrix element for the neutron-DM IND processes can be obtained from the proton-DM case using isospin symmetry, which relates the relevant hadronic matrix elements:
\begin{equation}
    \langle \pi^0 | (u_R d_R) d_R | n \rangle = \frac{1}{\sqrt{2}}\langle \pi^+ | (u_R d_R) d_R | p \rangle \,.
    \label{Eq:Isospin_Relation}
\end{equation}
The proton-DM IND cross sections will therefore be a factor of two larger than the corresponding neutron-DM cross sections (up to small isospin-violating corrections).

We take the values of the form factors from lattice QCD results~\cite{Aoki:2017puj} and from light cone sum rule (LCSR) calculations~\cite{Haisch:2021hvj} in $q^2$ regions where lattice results are unavailable. 

\begin{figure}
    \centering
    \includegraphics[width=\linewidth]{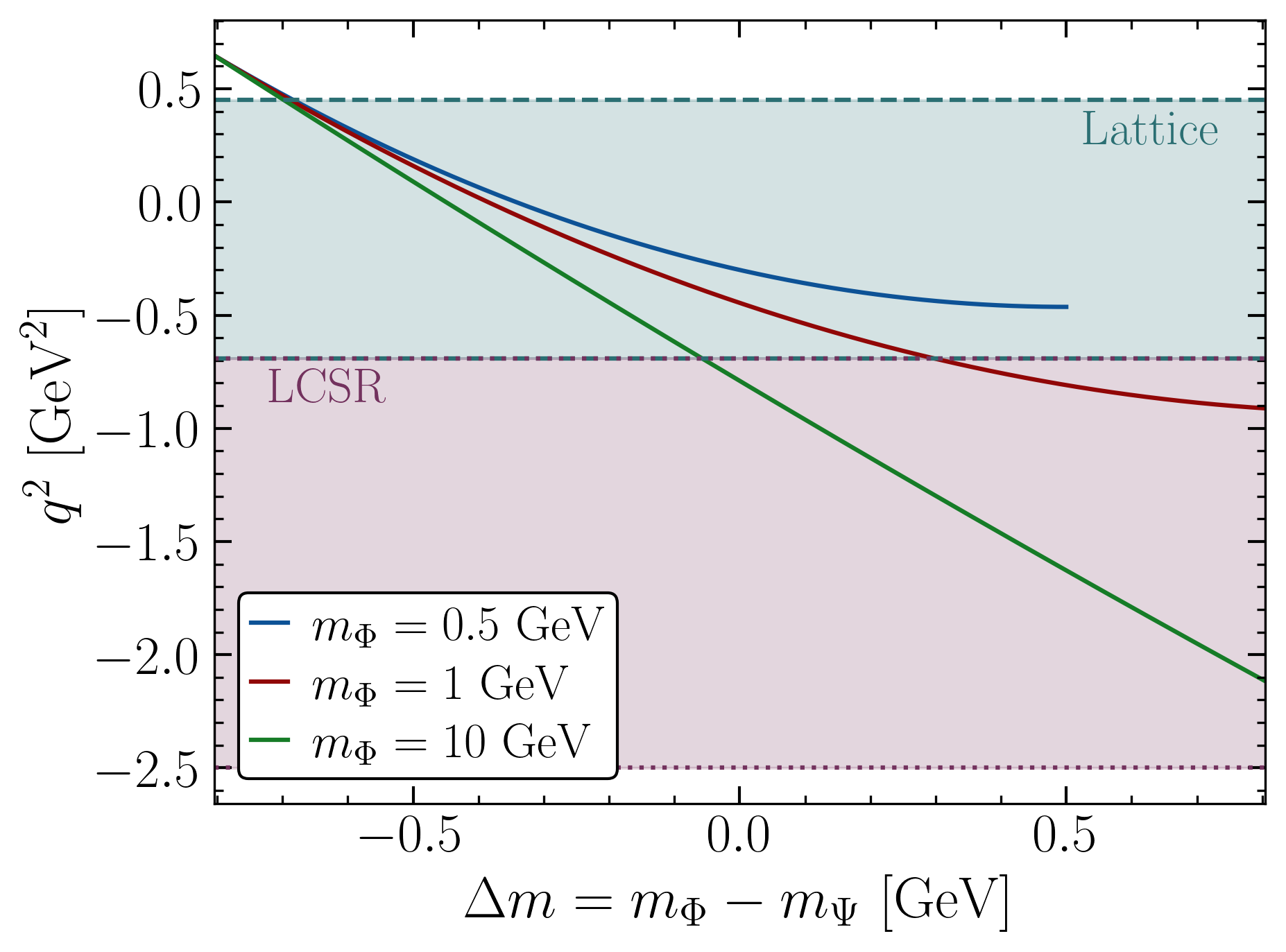}
    \caption{Momentum transfer, $q^2$, as a function of the mass splitting, $\Delta m$, for the neutron-$\Phi$ IND process. The upper shaded region represents the range of $q^2$ for which lattice QCD~\cite{Aoki:2017puj} form factors are used, the lower corresponds to LCSR~\cite{Haisch:2021hvj}.}
    \label{Fig:q2_Fn_DeltaM}
\end{figure}

Since the incoming DM and nucleon velocities are non-relativistic, $q^2$ is driven by the difference in the masses of the initial and final state particles. Figure~\ref{Fig:q2_Fn_DeltaM} shows the dependence of the momentum transfer on the mass splitting for the neutron-$\Phi$ IND process, $n + \Phi \rightarrow \bar{\Psi} + \pi^0$, for three benchmark values of the mass of the incoming DM particle, $m_{\Phi}$. The plot covers the full range of $\Delta m$ values consistent with Eq.~\eqref{Eq:Phi_Mass_Splitting_Requirements}. Note, the blue curve (corresponding to $m_{\Phi} = 0.5~\rm{GeV}$) is restricted to $\Delta m < 0.5~\rm{GeV}$, since $\Delta m \equiv m_\Phi - m_\Psi \leq m_\Phi$. The shaded regions denote the $q^2$ ranges over which lattice ($-0.65\,\mathrm{GeV}^2\leq  q^2 \leq 0.45\,\mathrm{GeV}^2$)~\cite{Aoki:2017puj} and LCSR ($-2.5\,\mathrm{GeV}^2 \leq q^2 < -0.65\,\mathrm{GeV}^2$)~\cite{Haisch:2021hvj} form factor values are used. For $\Delta m \lesssim -0.7~\rm{GeV}$, the value of $q^2$ falls outside these ranges and there are no available form factors. (For the neutron-$\Psi$ process this occurs for $\Delta m \gtrsim 0.7~\rm{GeV}$.) As such, we exclude this small range of $\Delta m$ values from our analysis. 

The general expression for the IND differential cross sections is given by
\begin{align}
    \frac{d \sigma_{\rm IND}}{dE_f} = \frac{|M|^2_{iN}}{32\pi m_N \left(E_i^2-m_i^2\right)} \, , 
    \label{Eq:Diff_Cross_Section}
\end{align}
where $i$ and $f$ denote the initial and final DM states, respectively. The velocity-averaged cross section is thus
\begin{align}
    \langle \sigma v \rangle_{\rm IND} =\int dE_f \int d^3 \mathbf{v} \, \frac{d \sigma_{\rm IND}}{dE_f} v f_{\rm Earth}(\mathbf{v}) \, ,
    \label{Eq:Vel_Avg_Cross_Section}
\end{align}
where $f_{\rm Earth}(\mathbf{v})$ is the DM velocity distribution boosted into the reference frame of the Earth. We take this to be a truncated Maxwellian distribution given by
\begin{align}
    f_{\rm Earth}(\mathbf{v})=\frac{1}{N(v_{0})} \exp & \left(-\frac{\left(\mathbf{v}+\mathbf{V}_{e}(t)\right)^{2}}{v_{0}^{2}}\right) \label{Eq:DM_Vel_Dist} \\
    & \times \Theta\left(v_{\rm esc}-\left|\mathbf{v}+\mathbf{V}_{e}(t)\right|\right) \, , \notag
\end{align}
where $N$ is a normalisation factor,
\begin{align}
    N(v_{0})=\pi^{3 / 2} v_{0}^{3}&\left[\operatorname{erf}\left(\frac{v_{\rm esc }}{v_{0}}\right) \right. \notag \\
    &\left. -\frac{2}{\sqrt{\pi}} \frac{v_{\rm esc}}{v_{0}} \exp \left(-\frac{v_{\rm esc}^{2}}{v_{0}^{2}}\right)\right]\, .
    \label{Eq:DM_Vel_Dist_Norm}
\end{align}
Adopting the recommended values from~\cite{Baxter:2021pqo}, we take the velocity dispersion to be $\sigma_0 = v_0/\sqrt{2}$ with $v_0 = 238~\rm{km/s}$, the escape velocity to be $v_{\rm esc} = 544~\rm{km/s}$, and the velocity of the Earth, $\mathbf{V}_e$, to simply be the velocity of the Sun, which is $|\mathbf{V}_e| = 250.2~\rm{km/s}$~\cite{Baxter:2021pqo}. We neglect the orbital velocity of the Earth around the Sun, as it is an order of magnitude smaller than the Sun's velocity through the Milky Way.

\begin{figure}
    \centering
    \includegraphics[width=\linewidth]{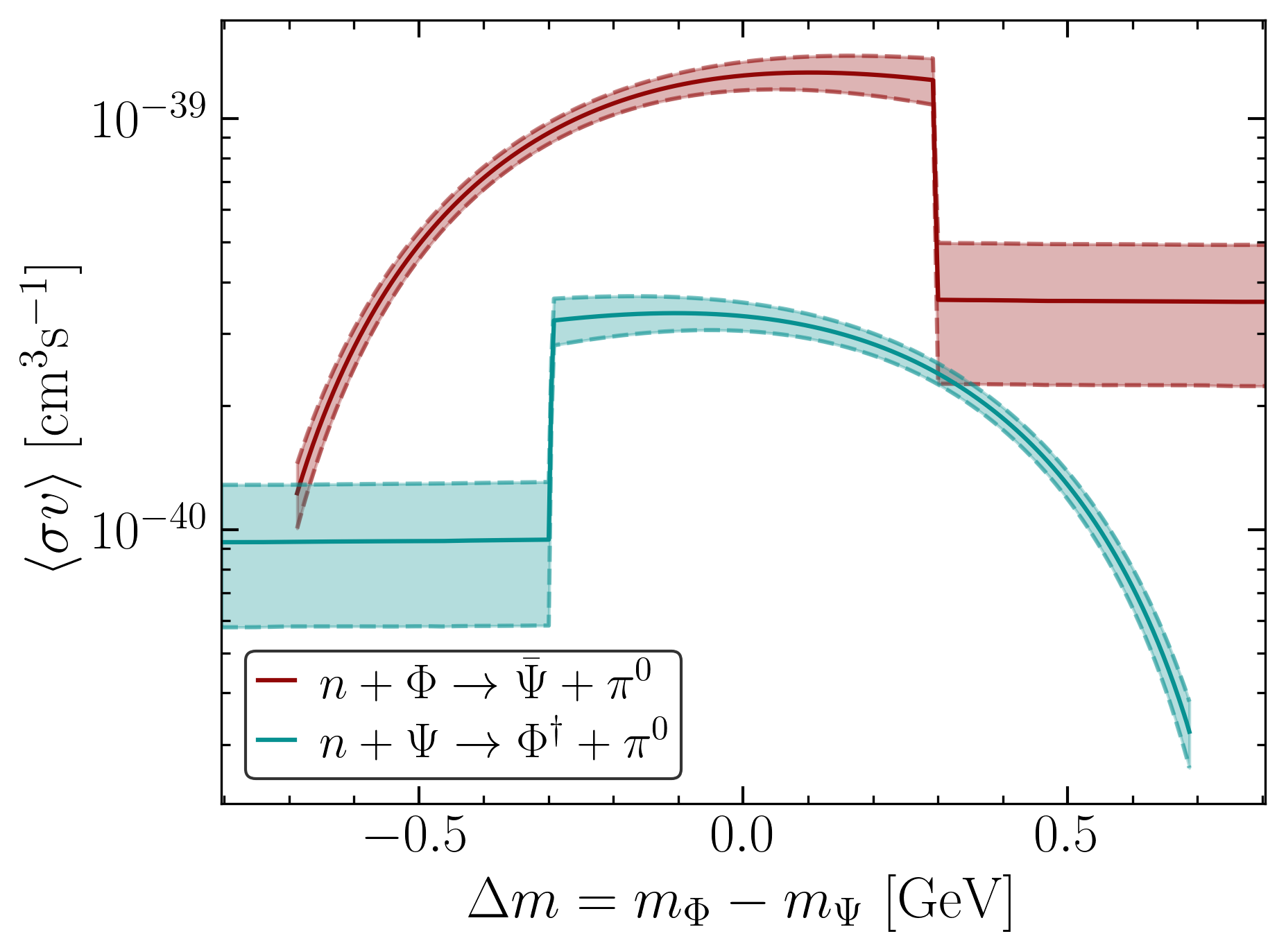}
    \caption{Velocity-averaged cross sections for the neutron-DM IND processes as a function of the DM mass splitting. The mass of the initial state DM is set to $1~\rm{GeV}$ and a benchmark value of $\Lambda = 1~\rm{TeV}$ is used. The discontinuities occur at the transition between lattice and LCSR form factors.}
    \label{fig:OneTeV_Cross_Section_Examples}
\end{figure}

Figure~\ref{fig:OneTeV_Cross_Section_Examples} shows the velocity-averaged cross sections for both neutron-DM IND processes as a function of the DM mass splitting. For each curve, the mass of the incoming DM particle is chosen to be $1~\rm{GeV}$ and a benchmark value of $\Lambda = 1~\rm{TeV}$ is used. The two curves are mirrored about $\Delta m = 0~\rm{GeV}$ due to the interchange of the $\Phi$ and $\Psi$ particles between the two processes and the definition of the mass splitting, $\Delta m = m_{\Phi} - m_{\Psi}$. Notice that the neutron-$\Psi$ IND cross section is smaller than the neutron-$\Phi$ cross section. While the matrix element takes the same form for both processes, the difference arises due to the fermion $\Psi$ being in either the initial or final state. As previously discussed, the proton-DM IND cross sections are a factor of two larger than the neutron cross sections due to the relation in Eq.~\eqref{Eq:Isospin_Relation}.

The discontinuities in Fig.~\ref{fig:OneTeV_Cross_Section_Examples} correspond to the transition from lattice to LCSR form factors at $q^2 = -0.65$ (as seen in Fig.~\ref{Fig:q2_Fn_DeltaM}). The error bands show the quoted uncertainties for the lattice~\cite{Aoki:2017puj} and LCSR~\cite{Haisch:2021hvj} form factors; the regions with larger error bands correspond to the LCSR results. Note that the uncertainties are too small to bridge the discontinuity in the transition region, suggesting that one, or both, of these calculations have underestimated their uncertainties.

\section{Reinterpretation of the Super-K nucleon decay analyses}
\label{Sec:Nuc_Decay_Reinterp}

As their name suggests, the pion-producing IND events have near-identical experimental signatures to the nucleon decay channels $n\rightarrow \pi^0 \nu$ and $p\rightarrow \pi^+ \nu$. The best constraints on these decays come from the Super-K analyses performed in~\cite{Super-Kamiokande:2013rwg}, which searched for single neutral and charged pions with momenta $p_\pi < 1$\,GeV. Here, we repurpose these data to place bounds on the scale of the neutron portal operator, $\Lambda$.

``Ordinary" nucleon decay channels produce mono-energetic pions with energy fixed by the two-body decay kinematics, up to corrections from Fermi motion. Therefore, the signal typically appears in only a few momentum bins around $p_\pi = 450$ MeV. IND processes are also effectively mono-energetic, due to the non-relativistic relative velocity of the incoming DM (though they are also affected by Fermi motion, which we discuss later). However, the pion momentum is now fixed by the mass splitting of the dark sector particles (given a fixed initial state DM mass). For example, for the nucleon-$\Phi$ IND case, the pion momentum is given by
\begin{equation}
    |\vec{p}_{\pi}|^2 \approx \left( \frac{(m_N + m_{\Phi})^2 + m_{\pi}^2 - (m_{\Phi} - \Delta m)^2}{2(m_N + m_{\Phi})} \right)^2 - m_{\pi}^2 \, .
\end{equation}
We therefore perform a series of single-bin analyses, setting bounds on $\Lambda$ for each value of $\Delta m$ and for fixed values of the DM mass.
 
For a given exposure, the expected number of IND events at Super-K, $N_{\rm{IND}}$, is given by:
\begin{align}
    N_{\rm{IND}} = n_{\rm{DM}} \times \langle \sigma v \rangle_{\rm{IND}} \times \epsilon \times N_{\rm{t}} \times t_{\rm{exp}} \, ,
    \label{Eq:Num_Exp_Events}
\end{align}
where $n_{\rm{DM}}$ is the local number density of DM, $\langle \sigma v \rangle$ is the velocity-averaged cross section (from Eq.~\eqref{Eq:Vel_Avg_Cross_Section}), $\epsilon$ is the signal efficiency, $N_{\rm{t}}$ is the number of targets, and $t_{\rm{exp}}$ is the exposure time. The analysis in~\cite{Super-Kamiokande:2013rwg} used data corresponding to an exposure of $172.8~\rm{kton}\cdot\rm{yrs}$. 

In our analysis, for each nucleon-DM IND process, we assume that the incoming DM particle (either $\Phi$ or $\Psi$) accounts for $100\%$ of the galactic DM. The results we obtain can be reinterpreted for alternative scenarios in which the DM is a mixture of the two states. We take the local DM density to be $\rho_{\rm{DM}} = 0.4~\rm{GeV/cm}^3$. Note that the bounds we obtain on the scale of the effective operator, $\Lambda$, scale only as the sixth root of the signal efficiency, $\epsilon$. In the following section, we make conservative choices to approximate the signal efficiency for each of the nucleon-DM IND processes.

\subsection{Neutron-DM IND efficiency}
\label{Subsec:Eff_Neut_Pi}

The neutron-DM IND process produces a single neutral pion in the final state. The dominant background at Super-K is therefore the neutral current process $\nu + n \rightarrow \nu + \pi^0$, where the incoming $\nu$ is an atmospheric neutrino. The Super-K $n\to\pi^0\nu$ analysis in~\cite{Super-Kamiokande:2013rwg} imposed the following selection cuts:
\begin{enumerate}
    \item There must be exactly two showering (i.e. electron-like) Cherenkov rings (from the $\pi^0 \to \gamma\gamma$ decay).
    \item The reconstructed invariant mass of the pion must be between $85~\rm{MeV}$ and $185~\rm{MeV}$.
    \item The reconstructed momentum of the pion must be less than $1~\rm{GeV}$.
\end{enumerate}
The efficiency curve for these selection cuts for the background process $\nu +n \to \nu + \pi^0$ is presented in~\cite{Super-Kamiokande:2013rwg} as a function of the pion momenta. Since the experimental signature is identical for both the signal and background processes (a single $\pi^0$), we can directly apply this efficiency curve to our neutron-IND signal process.

For certain values of the mass splitting $\Delta m$, the momentum of the pion produced by the IND process can exceed $1~\rm{GeV}$. Therefore, the selection cut on the pion momentum restricts the region of parameter space in which the Super-K analysis can be used to set constraints. A dedicated experimental analysis, extending to larger pion momenta, would be able to constrain additional parameter space at larger $|\Delta m|$.

In addition to the above selection cuts, the analysis in \cite{Super-Kamiokande:2013rwg} requires that events be fully-contained inside Super-K's Fiducial Volume (FV). For neutron-DM IND processes, an event is deemed fully-contained if the photons produced by the decay of the $\pi^0$ fully shower within the FV\footnote{The distance the neutral pion travels before decaying is negligible due to its extremely short lifetime.}. The maximum shower length of a photon is given by
\begin{equation}
    L_{\rm{max}} = X_0 \left( \ln{\frac{E_{\rm{max}}}{E_c}} + 0.5\right) \, ,
\end{equation}
where, for water, $X_0 = 36.17~\rm{cm}$, $E_c = 78.33~\rm{MeV}$, and $E_{\rm{max}}$ is the energy of the showering photon~\cite{ParticleDataGroup:2024cfk}. We implement this cut by defining an effective, fully-contained FV, defined by shrinking the FV in all dimensions by $L_{\rm{max}}$. To ensure this is conservative, $L_{\rm{max}}$ is calculated using the maximum possible photon energy (in the lab frame) for a given DM mass and mass splitting.

\subsection{Proton-DM IND efficiency}
\label{Subsec:Eff_Charged_Pi}

Next, we consider the proton-DM IND process, which has a single charged pion in the final state. Since Super-K can not distinguish between muons and charged pions, the dominant background is atmospheric neutrino charged-current quasi-elastic interactions of the form $\bar{\nu}_{\mu} + p \rightarrow \mu^+ + n$. The selection cuts used in the $p \rightarrow \pi^+ \nu$ analysis in~\cite{Super-Kamiokande:2013rwg} are:
\begin{enumerate}
    \item There must be exactly one non-showering (i.e. muon-like) Cherenkov ring.
    \item There is either zero or one electron-like rings. (Either one Michel electron from the decay of the muon or no electron if the pion is absorbed while travelling through the water).
    \item The reconstructed momentum of the pion must be less than $1~\rm{GeV}$.
\end{enumerate}
Ref.~\cite{Super-Kamiokande:2013rwg} provides efficiency curves for the background process $\bar{\nu}_{\mu} + p \rightarrow \mu^+ + n$ as a function of the muon momentum, with separate curves for zero and one electron events.\footnote{These are the curves labelled as muon-origin in Fig. 1 of \cite{Super-Kamiokande:2013rwg}.} While these efficiency curves can also be used to determine the selection efficiency for our signal process, we first need to determine the fraction of zero and one electron events. However, since the two efficiency curves differ by at most 20\%, we instead apply the one electron curve to all events for simplicity, which gives a conservative estimate of the signal efficiency.

We again have the additional requirement that the signal events be fully-contained within the FV. A fully-contained event requires all energy deposition to occur within the FV. For proton-DM IND events, we require the decay of the charged pion, the decay of the subsequent muon, and the energy loss of the Michel electron to all occur within the FV. Assuming that the charged pions lose kinetic energy at a rate of approximately $0.002~\rm{GeV/cm}$~\cite{ParticleDataGroup:2024cfk}, their stopping length is $L_{\rm{stop}} = T_{\pi} / (0.002~\rm{GeV/cm})$, where $T_{\pi}$ is the kinetic energy of the pion. The charged pion then decays at rest, producing a muon which is below Cherenkov threshold. Assuming the same kinetic energy loss as the charged pion, this muon travels a distance of approximately $2~\rm{cm}$ before coming to rest. The subsequent Michel electron will travel a similar distance before falling below Cherenkov threshold due to energy losses. To determine the fully-contained FV, we reduce the FV in all dimensions by the sum of these three distances (which is dominated by $L_{\rm{stop}})$.

Finally, unlike in the neutron-DM IND analysis, we need to separately take into account the effects of intranuclear pion interactions. This is because the selection efficiency curve we adopt from Ref.~\cite{Super-Kamiokande:2013rwg} is for muon-origin events, where there are no such effects. As the pion emerges from the nucleus it may be scattered, absorbed, or undergo charge exchange, altering the kinematics of the outgoing pion. Instead of performing a dedicated Monte Carlo simulation, we adopt the conservative approach of including only the fraction of pion events that do not interact with the nucleus. This underestimates the true signal efficiency, resulting in slightly weaker bounds. We take the fraction of non-interacting pions from Fig. $2$ of~\cite{Super-Kamiokande:2013rwg}, noting that while this applies to neutral pions, the charged pion case is stated to be ``similar". To account for possible differences, we normalise the total proton-DM IND signal efficiency such that it agrees with the signal efficiency quoted in Ref.~\cite{Super-Kamiokande:2013rwg} at the pion momentum for $p \rightarrow \pi^+ \nu$.

\subsection{Smearing of the pion momentum}
\label{Subsec:Pion_Smearing_Effects}

For both nucleon-DM IND and ordinary nucleon decays, the momentum of the outgoing pion is smeared due to the Fermi motion of the nucleons. The smearing effects of Fermi motion are of order $100~\rm{MeV}$, which is also the bin width used in the Super-K analysis. Therefore, this effect can smear the events across multiple momentum bins. To simplify the analysis, we simply double the momentum bin widths to $200~\rm{MeV}$ and perform the analysis for values of $\Delta m$ which produce pions with a momentum corresponding to the central value of each bin. This greatly reduces the likelihood of events being shifted into an adjacent bin, allowing us to neglect Fermi motion and perform a single-bin analysis.

\subsection{Statistical analysis}
\label{Sec:Stat_Analysis}

We now reinterpret the data from Fig.~3 of~\cite{Super-Kamiokande:2013rwg} to obtain $95\%$\,CL upper limits on the scale of the interaction, $\Lambda$, for both the $\Phi$ and $\Psi$ DM cases. We use a profile likelihood ratio approach, employing the $q_{\mu}$ test statistic defined in~\cite{Cowan:2010js} and assume that it is distributed according to Wilks' theorem. The smallest number of events in a pion momentum bin is approximately $10$, which is expected to be sufficient for Wilks' theorem to provide a good approximation. 

We present a single bound on $\Lambda$, combining both the neutron-DM and proton-DM IND processes via a joint likelihood function\footnote{Note that the proton- and neutron-IND processes are both kinematically allowed in almost all of the parameter space --- except in the tiny region at the edge of the parameter space where $m_p - m_{\pi^+} < \Delta m < m_n - m_{\pi^0}$.}. However, we find that the neutron-DM process is more sensitive due to a larger background for the proton-DM process.

To account for systematics, we introduce two nuisance parameters which normalise the expected number of background and signal events. We assume these parameters are uncorrelated and Gaussian distributed, with a mean of $1$ and a standard deviation of $0.3$ (based on the systematic uncertainties quoted in~\cite{Super-Kamiokande:2013rwg}). The dominant systematic is the theoretical uncertainty in modelling the pion-nucleus interactions. The theoretical uncertainty due to the form factors is not included in the statistical analysis. Instead, we separately show the impact of these uncertainties on our results in the figures in Sec.~\ref{Sec:Discussion}.

Additionally, we perform projections for the future Hyper-K detector, assuming an exposure of 3.8 megaton-years (20 years exposure). This is approximately $22$ times the Super-K exposure used in~\cite{Super-Kamiokande:2013rwg}. We assume that the expected background at  Hyper-K is the same as at Super-K, scaled up by the increase in exposure, and that all systematic uncertainties can be reduced by 50\% (i.e. the standard deviations for the nuisance parameters are halved from $0.3$ to $0.15$). The fully-contained FV cuts are recalculated for Hyper-K, with the remaining efficiencies assumed to be the same as Super-K.

\section{Results \& Discussion}
\label{Sec:Discussion}

\begin{figure}
  \centering
  \includegraphics[width=0.49\textwidth]{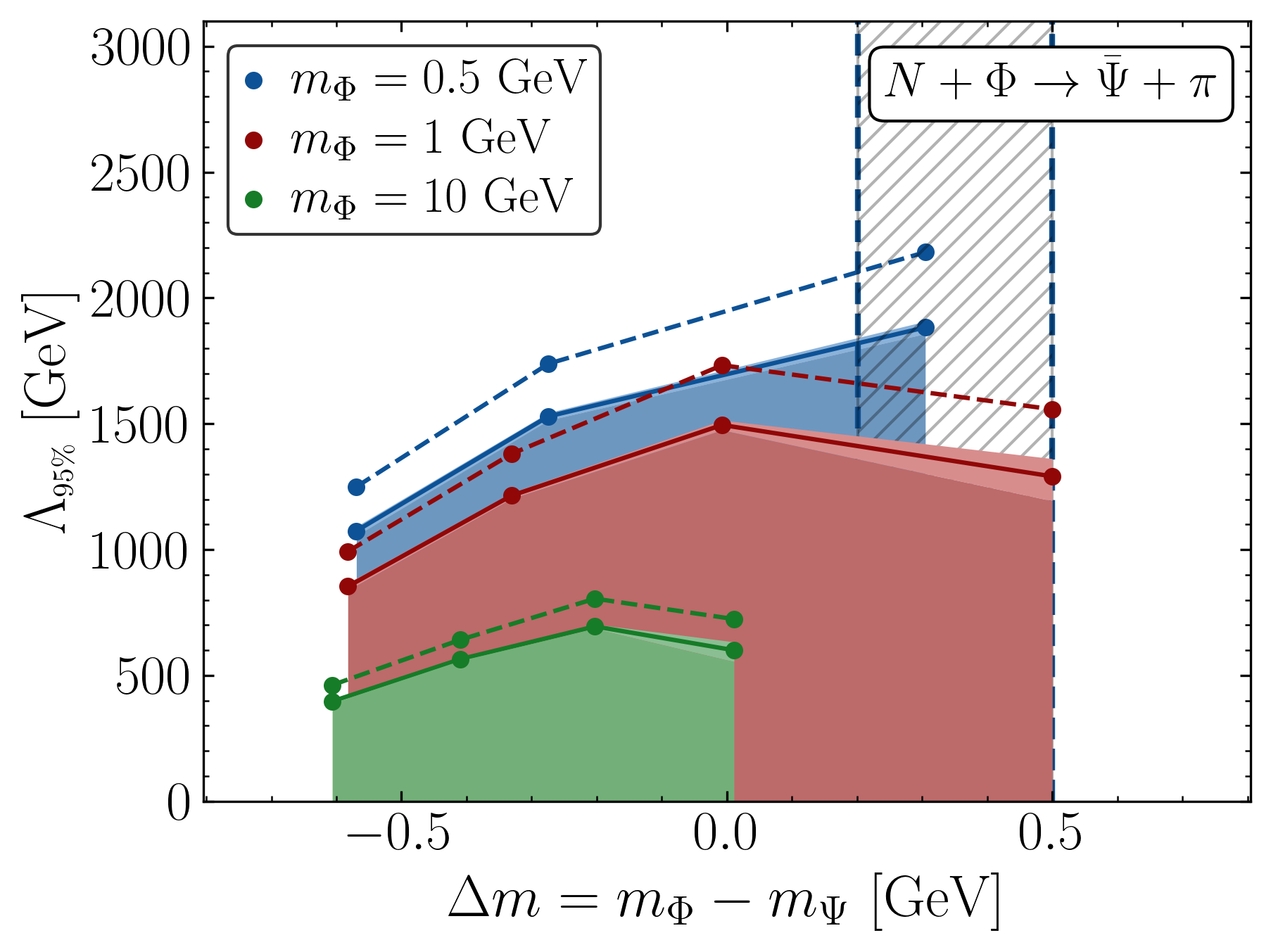}
  \hfill
  \includegraphics[width=0.49\textwidth]{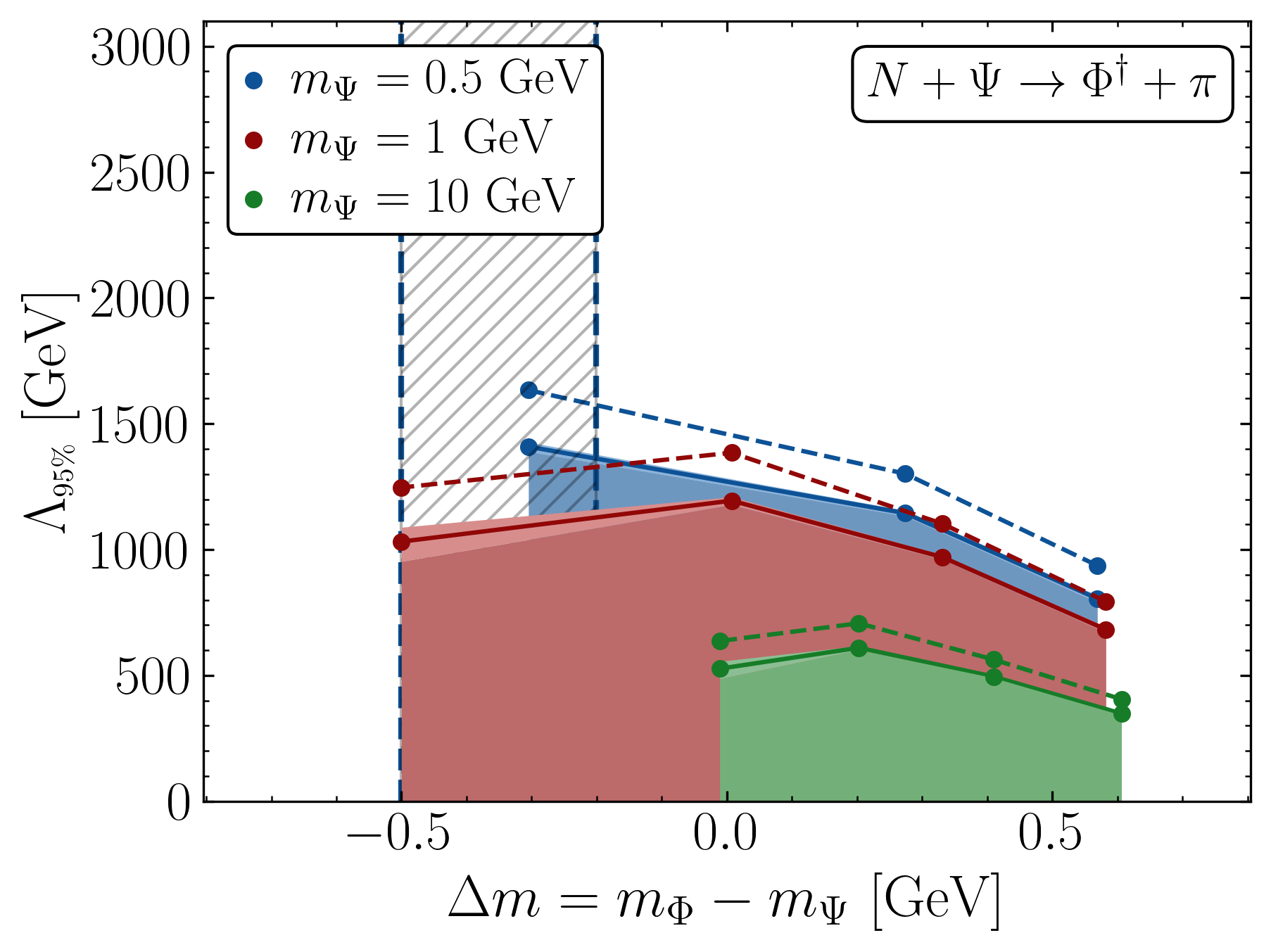}
  \caption{Bounds on the scale of the dimension-$7$ neutron portal operator at $95\%$\,CL from nucleon-$\Phi$ (top) and nucleon-$\Psi$ (bottom) IND. The solid curves use existing data from Super-K, while the dashed curves show projections for Hyper-K assuming a $50\%$ improvement of the systematics. In each plot, the incoming DM particle (either $\Phi$ or $\Psi$) is assumed to comprise $100\%$ of the galactic DM. The small lighter shaded regions represent the uncertainty due to the matrix element form factors. The hatched region shows where the direct proton decay constraints apply for a DM mass of $0.5~\rm{GeV}$ (blue curves).}
  \label{Fig:Nucloen_Lambda_Bound}
\end{figure}

Figure~\ref{Fig:Nucloen_Lambda_Bound} shows our $95\%$\,CL lower bounds on the scale $\Lambda$ as a function of the DM mass splitting and for three benchmark values of the incoming DM mass: $0.5, 1, \rm{and}~10~\rm{GeV}$. The shaded regions correspond to excluded values of $\Lambda$ and the small lighter shaded regions illustrate the impact of the uncertainties on the form factors. The uncertainty on the form factor has a minimal effect due to $\Lambda$ scaling as the sixth root of the cross section. As discussed in Sec.~\ref{Sec:IND_cross_section}, the nucleon-$\Phi$ cross section is larger than the nucleon$-\Psi$ cross section by a factor of approximately 4. This leads to the stronger bound on $\Lambda$ for the nucleon-$\Phi$ case.

We note that our bounds do not cover all kinematically allowed values of $\Delta m$. This is due to two limitations of the analysis:
\begin{enumerate}
    \item The upper limit on the pion momentum of 1 GeV, imposed by Super-K, limits the coverage of the curves to the right (left) of the upper (lower) plot of Fig.~\ref{Fig:Nucloen_Lambda_Bound}.
    \item The limited range of form factor values restricts the reach to the left (right) of the upper (lower) plots. For example, in the $\Phi$ case, no form factor values are available for $\Delta m$ values below $-0.7~\rm{GeV}$ (as seen in Fig.~\ref{Fig:q2_Fn_DeltaM}).
\end{enumerate}

Note that the $\Delta m$ ranges of the blue curves (corresponding to an incoming DM mass of $0.5~\rm{GeV}$) are further reduced due to the definition of $\Delta m$, which enforces $-m_\Psi \leq \Delta m \leq m_\Phi$. This gives the additional requirement $\Delta m < 0.5$\,GeV (nucleon-$\Phi$) and $\Delta m > -0.5$\,GeV (nucleon-$\Psi$). This is only relevant when the mass of the incoming DM particle becomes of order $\Delta m$ and so does not affect the other curves in Fig.~\ref{Fig:Nucloen_Lambda_Bound}. The hatched region illustrates the $\Delta m$ values for which proton decay can occur (blue curve only). This region is already strongly constrained by the ordinary proton decay analysis~\cite{Super-Kamiokande:2013rwg}.

Included in Fig.~\ref{Fig:Nucloen_Lambda_Bound} are the projections for Hyper-K (dashed curves), which show a minor improvement of about $15\%$ over the Super-K results. This should be compared to the naive expectation of $\sim30\%$ due to the increase in exposure. This indicates that despite the assumed $50\%$ improvement in the systematics (relative to Super-K), the projected sensitivity remains systematics limited. We note that without the reduced systematics, the projections show almost no improvement. As such, a dedicated experimental analysis with reduced systematics (particularly in the modelling of the pion-nucleus interactions) will be required if Hyper-K is to improve upon our bounds.

Finally, in the appendices, we also provide model-independent upper bounds on the velocity-averaged cross section for each of the INDs,
and reinterpret our bounds to constrain the Hylogenesis scenario of Ref.~\cite{Davoudiasl:2011fj}.

\section{Conclusion}
\label{Sec:Concl}

Neutron portal operators provide a well-motivated connection between the dark and visible sectors in ADM models where DM carries baryon number. In this paper, we studied the dimension-$7$ neutron portal operator that couples $udd$ to a dark fermion together with a dark scalar. This permits SM baryon-number violating processes in the form of induced nucleon decays, which can mimic ``ordinary" nucleon decays inside a detector. These signals are experimentally accessible to nucleon decay searches such as those performed by Super-K.

In this work, we reinterpreted the Super-K nucleon-decay searches for $n\rightarrow \pi^0\nu$ and $p\rightarrow \pi^+ \nu$ to place limits on the DM-induced counterparts, obtaining model-independent bounds on the scale of the neutron portal operator. For a Super-K exposure of 172.8 kiloton-years, we exclude $\Lambda \lesssim \mathcal{O}({\rm 1~TeV})$ for GeV-mass DM. We project that next-generation nucleon decay searches at Hyper-K, with $\sim$20-times greater exposure, would improve the sensitivity to $\Lambda$ by a modest $15\%$, assuming a 50\% reduction in systematic uncertainties compared to the Super-K analysis. The present analysis is systematics limited, predominantly due to the uncertainty on the pion-nucleus interaction modelling and, as such, we do not expect any improvement from Hyper-K without a commensurate reduction in these uncertainties. This work highlights the importance of a dedicated experimental analysis for Hyper-K with reduced systematics, which would broaden the range of DM masses that can be constrained and enhance the exclusion power of future detectors.

\section*{Acknowledgements}
The authors would like to thank Stephan Meighen-Berger, Ryo Matsumoto, and John Gargalionis for informative discussions. This work was supported in part by the ARC Centre of Excellence for Dark Matter Particle Physics, CE200100008. P.C. was supported by the Australian Research Council Discovery Early Career Researcher Award DE210100446. M.B.G.V was supported by a University of Melbourne Research Scholarship.

\appendix

\section{Constraints on the Hylogenesis Model}
\label{Sec:Appdx_Hylogenesis}

\begin{figure}[t]
    \centering
    \includegraphics[width=\linewidth]{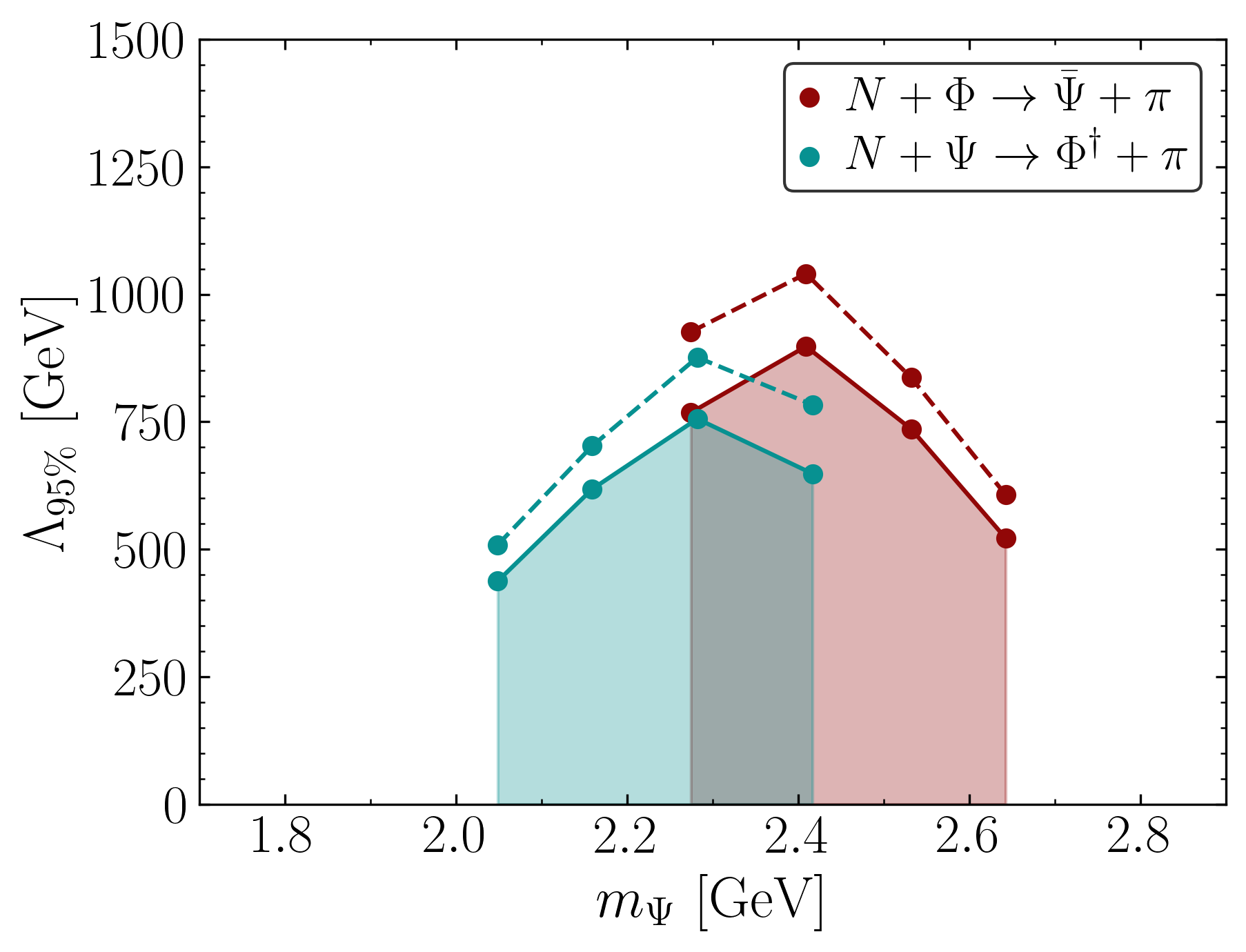}
    \caption{IND constraints on the EFT scale $\Lambda$ in Hylogenesis models (95\%\,CL), which satisfy the additional mass requirement $m_\Phi = 5m_p - m_\Psi$ with $m_\Psi = \{1.7, 2.9\}~\rm{GeV}$. These are shown for the nucleon-$\Phi$ (red) and nucleon-$\Psi$ (blue) IND processes.}
    \label{Hylo_Lambda_Bound}
\end{figure}

The results of this analysis can be directly applied to the Hylogenesis models presented in~\cite{Davoudiasl:2010am,Davoudiasl:2011fj}. These models impose that $n_B = n_\Phi = n_\Psi$, which then requires $m_\Psi + m_\Phi \approx 5m_p$ to ensure the correct DM relic abundance. This is an additional mass requirement, beyond those discussed in Sec.~\ref{Sec:EFT}. The combined mass requirements impose $m_\Psi = \{1.7, 2.9\}~\rm{GeV}$, with $m_\Phi = 5m_p - m_\Psi$~\cite{Davoudiasl:2011fj}. In Fig.~\ref{Hylo_Lambda_Bound}, we present bounds on $\Lambda$, as a function of $m_\Psi$, with these mass relations imposed. Included on this plot are the projections for Hyper-K (shown by the dashed lines) under the same assumptions discussed in Sec.~\ref{Sec:Nuc_Decay_Reinterp}.

\section{Constraints on the IND Cross Sections}
\label{Sec:Appdx_xsec}

\begin{figure*}[t]
    \centering
    \includegraphics[width=0.49\linewidth]{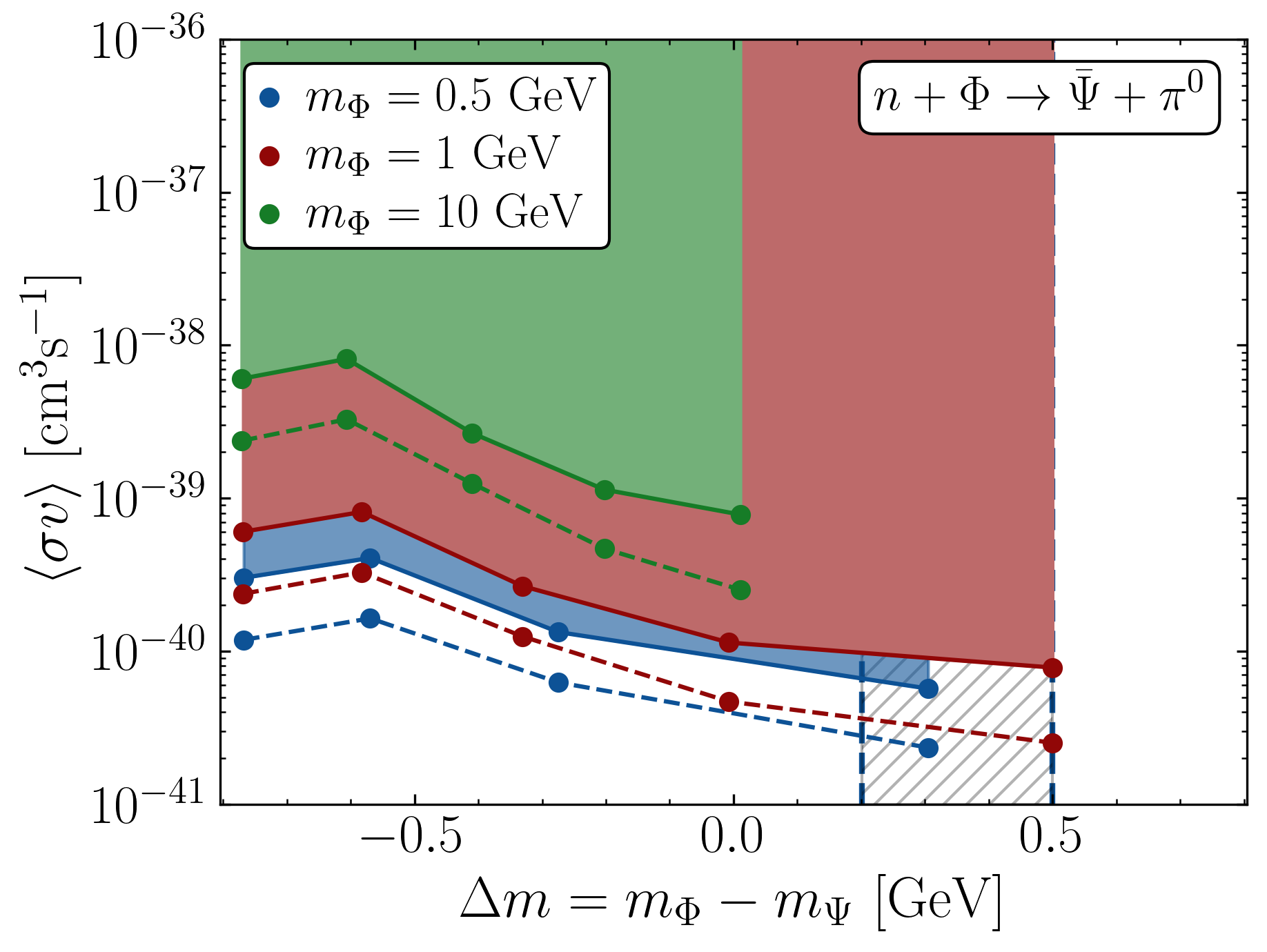}
    \hfill
    \includegraphics[width=0.49\linewidth]{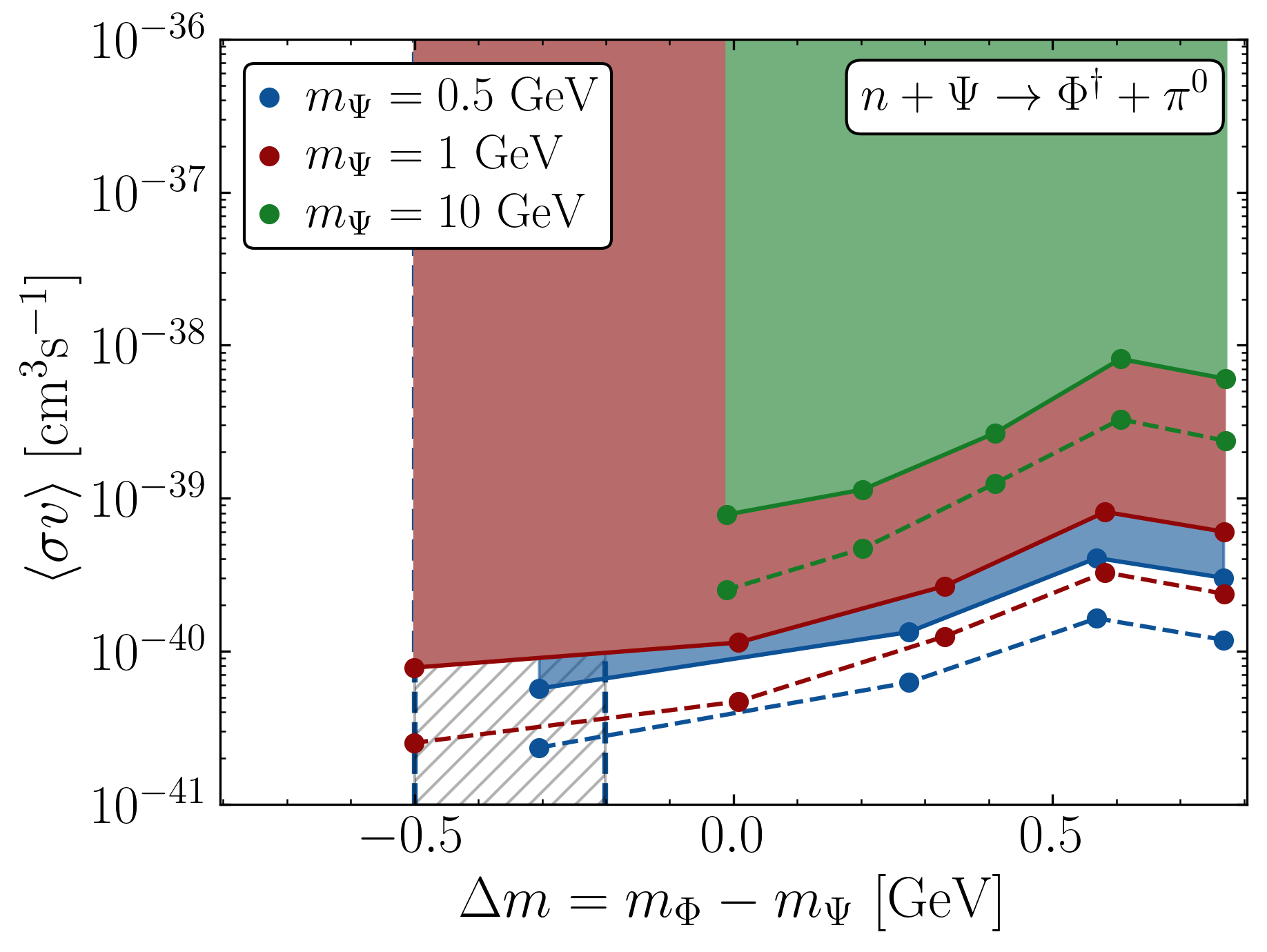}
    \\\vspace{0.5em}
    \includegraphics[width=0.49\linewidth]{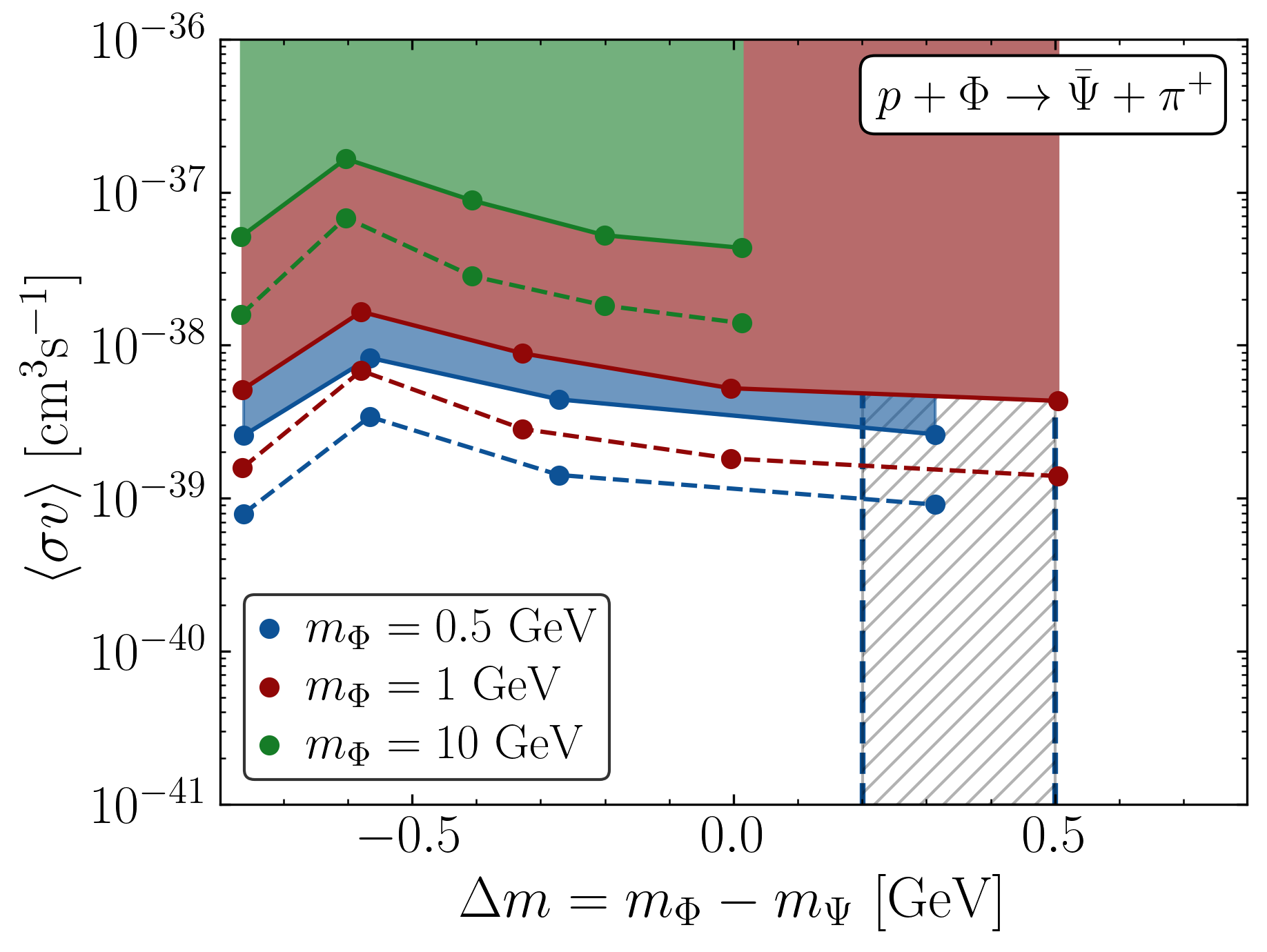}
    \hfill
    \includegraphics[width=0.49\linewidth]{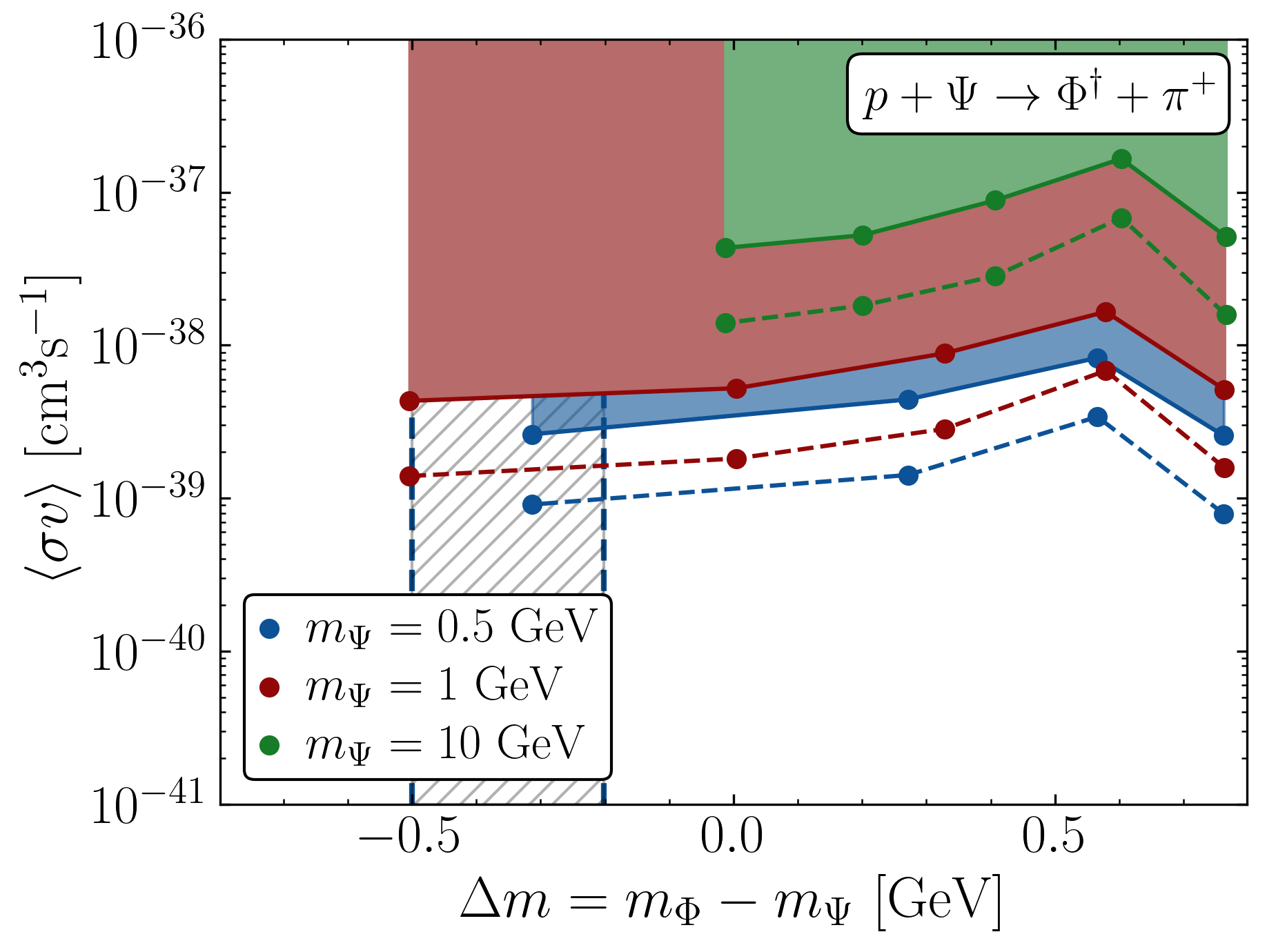}
  \caption{Upper bounds on the IND velocity-averaged cross sections from Super-K (solid) and projected bounds for Hyper-K (dashed). The top (bottom) row is for neutron-DM (proton-DM) IND processes and the left (right) column is for the nucleon-$\Phi$ (nucleon-$\Psi$) process. The hatched region shows where the direct proton decay constraints apply for a DM mass of $0.5~\rm{GeV}$ (blue curves).}
  \label{Fig:Cross_Section_Bound}
\end{figure*}

In this appendix, we provide model-independent upper bounds on the IND cross sections. These are shown in Fig.~\ref{Fig:Cross_Section_Bound} for each individual nucleon-DM IND process. The upper (lower) row corresponds to neutron-DM (proton-DM) and the left (right) column to IND processes with $\Phi$ ($\Psi$) in the initial state. Since these constraints do not depend on the form factors, which restricted the values of $q^2$, there is an additional $\Delta m$ point compared to the bounds on $\Lambda $ in Fig.~\ref{Fig:Nucloen_Lambda_Bound}.

\newpage

\bibliographystyle{apsrev4-2}
\bibliography{ref}

\end{document}